\documentclass[a4paper, 11pt]{article}

\usepackage{indentfirst}
\usepackage{epsfig}
\usepackage{psfrag}
\usepackage{graphicx}
\usepackage{overpic}
\usepackage{multirow}
\usepackage[dvips]{hyperref}
\usepackage[hang]{subfigure}
\usepackage{amsmath,amssymb, amsthm}
\usepackage[dvips]{color}
\usepackage[mathscr]{euscript}

\newtheorem{theorem}{Theorem}

\newcommand{\figref}[1]{\figurename~\ref{#1}}

\newcommand\bbR{\mathbb{R}}
\newcommand\bbN{\mathbb{N}}
\newcommand\bxi{\boldsymbol{\xi}}
\newcommand\bx{\boldsymbol{x}}
\newcommand\by{\boldsymbol{y}}
\newcommand\bp{\boldsymbol{p}}

\newcommand\bu{\boldsymbol{u}}
\newcommand\bv{\boldsymbol{p}}

\newcommand\bn{\boldsymbol{n}}

\newcommand\blambda{{\boldsymbol{\lambda}}}

\newcommand\dd{\,\mathrm{d}}
\newcommand\He{\mathit{He}}

\newcommand\bw{\boldsymbol{w}}
\newcommand\RM{{\cal R}_{M}}
\newcommand\cS{{\cal{S}}}
\newcommand\mT{\mathcal{T}}
\newcommand\mH{\mathcal{H}}

\newcommand\pd[2]{\dfrac{\partial {#1}}{\partial {#2}}}
\newcommand\opd[2]{\dfrac{\dd {#1}}{\dd {#2}}}

\newcommand\mN{{\mathcal N}}
\newcommand\rC[2]{{\rm{C}}_{{#1},{#2}}}

\newcommand\comment[1]{}

\graphicspath{{images/}}
{\theoremstyle{remark} \newtheorem{remark}{Remark}}

\title{Quantum Hydrodynamic Model by Moment Closure of Wigner
  Equation}

\author{Zhenning Cai\thanks{School of Mathematical Sciences, Peking
    University, Beijing, China, email: {\tt caizn@pku.edu.cn}.},~~
  Yuwei Fan\thanks{School of Mathematical Sciences, Peking University,
    Beijing, China, email: {\tt ywfan@pku.edu.cn}.},~~ Ruo
  Li\thanks{HEDPS \& CAPT, LMAM \& School of Mathematical Sciences,
    Peking University, Beijing, China, email: {\tt
      rli@math.pku.edu.cn}.},~~ Tiao Lu\thanks{HEDPS \& CAPT, LMAM \&
    School of Mathematical Sciences, Peking University, Beijing,
    China, email: {\tt tlu@math.pku.edu.cn}.},~~ Yanli
  Wang\thanks{HEDPS \& CAPT, School of Mathematical Sciences, Peking
    University, Beijing, China, email: {\tt wangyanliwyl@gmail.com}.}
}

\begin{document}
\maketitle
% vim: tw=70:spell
\begin{abstract}
  In this paper, we derive the quantum hydrodynamics models based on
  the moment closure of the Wigner equation. The moment expansion
  adopted is of the Grad type firstly proposed in \cite{Grad}. The
  Grad's moment method was originally developed for the Boltzmann
  equation. In \cite{Fan_new}, a regularization method for the Grad's
  moment system of the Boltzmann equation was proposed to achieve the
  globally hyperbolicity so that the local well-posedness of the
  moment system is attained. With the moment expansion of the Wigner
  function, the drift term in the Wigner equation has exactly the same
  moment representation as in the Boltzmann equation, thus the
  regularization in \cite{Fan_new} applies. The moment expansion of
  the nonlocal Wigner potential term in the Wigner equation is turned
  to be a linear source term, which can only induce very mild growth
  of the solution. As the result, the local well-posedness of the
  regularized moment system for the Wigner equation remains as for the
  Boltzmann equation.

\vspace*{4mm}
\end{abstract}

\section{Introduction}
The Wigner equation was proposed by Wigner in 1932 as a counterpart in
quantum mechanics of the Boltzmann equation
\cite{Wigner1932}. Different from the distribution function in the
Boltzmann equation, the Wigner function may take negative values, so
it is not a true probability distribution function. But it is called a
``quasi-probability distribution function" in the phase space because
it allows one to express quantum mechanical averages (related to
statistical moments) in a form which is very similar to that for
classical averages. Readers may refer to the review article by Hillery
{\it et al.} \cite{Hillery1984} for more details of the properties of
the Wigner function. It offers a convenient interpretation of the
ensemble in the form of a quasi-probability distribution in the phase
space defined by position $\bx$ and momentum $\bp$ and has been proved
to of great use not only as a calculation tools but can also provide
insights into the connections between classical and quantum mechanical
mechanics. Actually, Lions {\it et al.} have given a rigorous
mathematical proof that in the semiclassical limit, the Wigner
equation gives the Vlasov (or Liouville) equation in the phase space
\cite{Lions1993}.  As one of the most accurate equations in quantum
mechanics, the Wigner equation has many advantages over the
Schr\"odinger equation, the nonequilibrium Green method and the
density matrix method in simulating the carrier transport in
semiconductor devices because the description of boundary conditions
and collision operator for the semi-classical Boltzmann equation can
be extended to the Wigner equation due to their strong similarity
\cite{ferry_transport}. There has been an increasing interest in the
Wigner equation as the size of semiconductor devices goes into the
nano-scale under which the quantum effects becomes not negligible any
more \cite{itrs2009}. In some devices, the quantum effect even plays a
dominating role, e.g., the resonant tunneling diode (RTD) has been
extensively studied recently using the numerical methods based on the
Wigner equation \cite{Frensley1987}.  The numerical method for the
Wigner equations has attracted many researchers from different fields
\cite{ShLu09, GeKo05, dollfus09, KKFR89, JaBo04, KoNe06, zhao_05,
  Biegel1996}. The deterministic numerical methods have been
successfully used in simulating one-dimensional devices, but are not
expected to be directly used for multi-dimensional devices simulation
due to its formidable expense in memory storage and computational
time.  One practical approach to investigate a higher dimensional
devices where quantum effects are relevant is to use quantum
hydrodynamics models which are moment systems derived from the Wigner
equation.  Because the close connection of the Wigner equation and the
Boltzmann equation, many moment methods devised for the Boltzmann
equation have been extended to the Wigner equation, see
\cite{Gardner1994, Degond2003} and references therein. Equations
derived from the Wigner equation are called quantum drift-diffusion
equations, quantum Euler equations and quantum hydrodynamics equation,
and numerical simulations based on such moment equations are
extensively studied \cite{Zhou1992, Degond2005QETDD, Jungel1997,
  Tang2008}. In this paper, we will extended the moment method
recently proposed in \cite{Fan_new} for the Boltzmann equation to the
Wigner equation.

In 1940s, Grad \cite{Grad} proposed a moment method to approximate the
Boltzmann equation, and a 13-moment model is given as an extension of
the classic Euler equations. However, this model was quickly found to
be problematic \cite{Grad1952}. Its major deficiencies include the
appearance of subshocks in the structure of a strong shock wave and
the loss of global hyperbolicity. Later on, a number of
regularizations were raised to solve or alleviate these problems.
Levermore \cite{Levermore} introduced a promising way to achieve
global hyperbolicity, while 14 moments are needed and the explicit
expressions of the equations cannot be written. Jin and Slemrod
\cite{Jin} gave a regularization of the Burnett equations via
relaxation, which resulted in a set of equations containing the same
variables with Grad's 13-moment theory, and no subshocks appear in the
structure of shock waves. Struchtrup and Torrilhon
\cite{Struchtrup2003} regularized Grad's system by integrating the
moment method with Chapman-Enskog expansion, and the resulting system
is called as the R13 equations. And in \cite{Struchtrup2004},
Struchtrup improved the R13 equations by the ``order of magnitude''
method. The R13 system also removes the discontinuities in the shock
wave, and it extends the region of hyperbolicity considerably
\cite{TorrilhonRGD}. Recently, Torrilhon \cite{Torrilhon2010} tackled
the problem of hyperbolicity by introducing the Pearson-Type-IV
distributions, and the resulting system is proven to be able to gain a
much larger hyperbolicity region.

Due to the complexity of the explicit expressions, systems with large
number of moments are not investigated until recently. In
\cite{Torrilhon}, Torrilhon and his coworkers developed a software
named $ET_{XX}$ \cite{ETXX} which is able to generate moment systems
with almost any number of moments in one-dimensional space. And in
\cite{Weiss}, some numerical results for a shock tube were carried out
to show the behavior of characteristic waves in the extended
thermodynamics. A numerical method solving large moment systems was
proposed in \cite{NRxx}, and therein the regularization technique in
\cite{Struchtrup2003} was applied to general moment systems. In
\cite{NRxx_new}, the order of magnitude method was also integrated
into large moment systems. In \cite{Fan}, the authors focused on the
one-dimensional velocity space, and the characteristic polynomial of
the quasi-linear coefficient matrix is found to be very simple; thus a
brand-new regularized model with global hyperbolicity is proposed by
the correction of the characteristic speed. Such regularization is
extended to the multi-dimensional velocity space in \cite{Fan_new}.

It is clear that the moment expansion of the drift term for Boltzmann
equation can be extended to the Wigner equation. The resulting
convective terms in the moment system expanded from the drift term of
the Wigner equation has exactly the same format as that of the
Boltzmann equation. Thus the method of the hyperbolic regularization
in \cite{Fan_new} is applied to the Wigner equation, too, to achieve
the global hyperbolicity. As the difference of the Wigner equation
from the Boltzmann equation, the nonlocal Wigner potential term due to
the electric potential is also expanded using Hermite polynomial. It
is interesting for us to find that this term can be represented in the
moment expansion style with very compact expressions. The overall
formation of the nonlocal Wigner potential term in the moments is
turned into a linear source term, with compact sparse coefficient
matrix. Moreover, the contribution from the lower order moments in
this term is always on the higher order moments. Immediately, the
coefficient matrix in the linear source term is strictly lower
triangular, thus it is a nilponent matrix. Noticing that the moment
expansion of the relaxation scattering term produces a linear source
term providing an exponential decay of the high order moments. As a
result, the growth of the high order moments in time due to the source
term is essentially slower than exponential growth rate. This makes
that the derived moment system is formulated as a quasi-linear system,
plus a linear source term which can induce only very mild growth of
the high order moments. Since the convection term in the system is
guaranteed to be globally hyperbolic by the regularization, the local
well-posedness of the system is out of box.

The rest of this paper is arranged as follows: in Section 2 we present
the elementary formula of Wigner equation. The moment expansion of
Wigner equation is carried out in Section 3 and the system obtained is
closed by truncation of the expansion and regularized using method in
\cite{Fan_new} in Section 4 to achieve the final hyperbolic system. In
Section 5, we discuss the simple case in 1D space for better
understanding of the structure of the derived moment
system. Concluding remarks are in the last section.

%%% Local Variables: 
%%% mode: latex
%%% TeX-master: "article"
%%% End: 

% vim: tw=70:spell

\section{Wigner Equation with Smooth Potential}
We consider a one-particle system in a statistical mixture with $N$
states described by the wave functions $\Psi_{k} = \Psi_k (t,\bx)$,
each with a probability $P_k \geqslant 0 $, with
$k=1,\cdots,N_{\text{state}}$, which satisfies
$\sum_{k=1}^{N_{\text{state}}} P_k = 1$. The wave functions
$\Psi_k(t,\bx)$ of state $k$ obeys the Schr\"odinger equation
\begin{equation}
\label{eq:schrodinger}
i \hbar \pd{\Psi_{k}(t,\bx)}{t} = -\frac{\hbar^2}{2m}\nabla_{\bx}^2
		\Psi_{k}(t,\bx)
+ V(t,\bx) \Psi_{k}(t,\bx) , 
\end{equation}
where $i=\sqrt{-1}$, $\hbar$ is the reduced Planck constant, $m$ is the particle
effective mass assumed to a constant in this paper, and $V(t,\bx)$ is
the electric potential energy (which will be called potential for
short hereafter). We can also regard the one-particle system as a
many-particle system with $P_k$ interpreted as the percentage of the
particles occupying state $k$.  This is the general way the Wigner
equation used in modeling carrier transport in semiconductor devices.
We construct the Wigner function $f(t,\bx,\bp)$ which is a
quasi-probability density function in the phase space $(\bx,\bp)$, as
usual, by
\begin{equation}
  f(t,\bx,\bp) = 
  \frac{1}{(2\pi \hbar)^3} \sum_{k=1}^{N_{\text{state}}} P_k
  \int_{\bbR^3} \Psi_k^{*}(t, \bx + \by/2) \Psi_k(t,\bx-\by/2)
	e^{i\by\cdot \bp/\hbar} d^3 \by. 
\end{equation}
Using the Schr\"odinger equation \eqref{eq:schrodinger} and the
definition of the Wigner function given above, we can derive the
Wigner equation (or the quantum Vlasov equation) as been done in
\cite{Wigner1932, Hillery1984}
\begin{equation}
  \label{eq:vlasovBallistic}
  \pd{f}{t} + \frac{\bv}{m} \cdot \nabla_{\bx} f + \Theta[V] f
  = 0, 
  \quad \bx \in \bbR^3, \ \bp \in \bbR^3 ,
\end{equation}
where the nonlocal Wigner potential term $\Theta[V] f$ is a 
pseudo-differential operator. It is defined by
\begin{equation} 
\label{eq:ThetaV}
(\Theta[V]f)(t,\bx,\bv)=
\int_{\bbR^3}
V_w(t,\bx,\bv')
 f(t,\bx,\bv-\bv') \dd  \bp',
\end{equation}
where the Wigner potential is as
\begin{equation*}
V_{w}(t,\bx,\bv)=
\frac{-i}{(2\pi\hbar)^3 \hbar} \int_{\bbR^3}
\left[ V\left(t, \bx+\frac{\by}{2}\right)
		-V\left(t, \bx-\frac{\by}{2}\right) \right] 
e^{i \by\cdot \bv/\hbar 
}  \dd \by . 
\end{equation*}
For convenience, we simply take $m = 1$ later on. Similar to a
Vlasov-Poisson system, a Wigner-Poisson system could be also
considered to include the self-consistent electric field induced by
the redistribution of the electrons. In this paper, we focus on the
case that $V(t,\bx)$ is a given smooth function.  The pseudo-operator
$\Theta[V] f $ can be written into an equivalent form
\cite{Hillery1984}
\begin{equation}
\label{eq:ThetaVDiff}
(\Theta[V]f)(t,\bx,\bv)=
-\sum_\blambda
\frac{ (\hbar/2i) ^{ |\blambda| -1 }}{ \blambda ! }
\frac{ \partial ^{\blambda } V} 
{\partial \bx ^ {\blambda} } 
\frac{ \partial ^{\blambda } f} 
{\partial \bv ^ {\blambda} } ,
\end{equation} 
where $\blambda=(\lambda_1,\lambda_2,\lambda_3)$,
$|\blambda|=\sum_{i=1}^3 \lambda_i $, $\bx
^{\blambda} = \prod_{i=1}^3 x_i^{\lambda_i}$,  
\[
\dfrac{\partial^\blambda}{\partial \bx^\blambda} = \prod_{i=1}^3
\dfrac{\partial^{\lambda_i}}{\partial x_i^{\lambda_i}}, \qquad
\dfrac{\partial^\blambda}{\partial \bp^\blambda} = \prod_{i=1}^3
\dfrac{\partial^{\lambda_i}}{\partial p_i^{\lambda_i}},
\]
and the summation over $\blambda$ has to be extended over all
non-negative integer values of $\lambda_1,\lambda_2,\lambda_3$ for
which $\vert \blambda \vert$ is odd. In the semi-classical limit
$\hbar\rightarrow 0$, $\Theta[V]f$ converges to the usual operator
$-\nabla_{\bx}V \cdot \nabla_{\bp} f$. The equation
\eqref{eq:vlasovBallistic} describes particles movement without
scattering corresponding to ballistic transport. The scattering effect
may be considered by adding a scattering term to the right hand side of
\eqref{eq:vlasovBallistic}. So one obtains the following Wigner
equation with a scattering term:
\begin{equation}
  \label{eq:vlasov}
  \pd{f}{t} + \bv \cdot \nabla_{\bx} f + \Theta[V] f
  = \left. \pd{f}{t} \right \vert_{\mathrm{Scat}}, 
  \quad \bx \in \bbR^3, \ \bp \in \bbR^3. 
\end{equation}
The time-relaxation approximation of the scattering term
is often used for the Wigner equation \cite{Frensley1987}, and it has
the same form of the BGK scattering term used for the collision of
gas \cite{BGK}. It is easy to be study analytically and takes 
of the following simple form
\begin{equation}
\left. \pd{f}{t} \right \vert_{\mathrm{Scat}}
= \frac{ f_{\mathrm{eq}} - f } {\tau} ,
\end{equation}
where $\tau$ is the relaxation time and $f_{\mathrm{eq}}$ is the 
equilibrium distribution. For example, when the density of electrons
are not extremely high, we can assume that the equilibrium 
distribution is a Maxwellian distribution,
\begin{equation}
\label{eq:Maxwellian}
f_{\mathrm{eq}}(t,\bx,\bp) = 
\frac{\rho(t,\bx)}{ \left( 2\pi k_B T  \right)^{3/2}} 
\exp\left( - \frac{(\bp - \bu(t,x))^2}{2 k_B T(t,\bx)} \right)
\end{equation}
where $\rho(t,\bx)$ is the number density of particles at position
$\bx$, $k_B $ is the Boltzmann constant, $T(t,\bx)$ is the particle
temperature, and $\bu(t,\bx)$ is the average momentum of particles.
These macroscopic variables are related with the distribution function
as below:
\begin{equation}
\rho(t,\bx) = \int_{\bbR^3} f(t,\bx,\bv) \dd \bv , 
\end{equation}
\begin{equation}
\rho(t,\bx)\bu(t,\bx) = \int_{\bbR^3} \bv f(t,\bx,\bv) \dd \bv , 
\end{equation}
\begin{equation}
\label{eq:DefTemperature}
\rho(t, \bx) k_B T(t,\bx) = \int_{\bbR^3} |\bv - \bu|^2 f(t,\bx,\bv)
	\dd \bv . 
\end{equation}

The conservation of mass, momentum and total energy are all valid for
the Wigner equation. Multiplying the equation \eqref{eq:vlasov} by
$1$ and $\bv$, direct integration with $\bv$ and $\bx$ gives us
\begin{gather}
  \label{eq:mass_conservtion}
  \frac{\dd }{\dd t}\int_{\bbR^3\times
    \bbR^3}f(t,\bx,\bv)\dd \bx \dd \bv = 0,
  \quad  t
  \in \bbR^+, \\
\label{eq:momentun_conservation}
\frac{\dd }{\dd t}\int_{\bbR^3\times \bbR^3}\bv f(t,\bx,\bv)\dd \bx
\dd \bv = - \int_{\bbR^3} \rho(t,\bx) \pd{V(t,\bx)}{\bx} \dd \bx, \quad t \in
\bbR^+.
\end{gather} 
Multiplying the equation \eqref{eq:vlasov} by $|\bv|^2$ and
integrating by parts, we get the conservation of the total energy for
the system \eqref{eq:vlasov} and \eqref{eq:ThetaVDiff}:
\begin{equation}
  \label{eq:energy_conversvation}
  \frac{\dd }{\dd t}\int_{\bbR^3\times \bbR^3}f(t,\bx,\bv)|\bv|^2\dd
  \bx\dd \bv 
= -2\int_{\bbR^3}\rho(t,\bx) \bu(t,\bx)\cdot\nabla_{\bx}
  V(t, \bx)\dd \bx, \quad  t \in \bbR^+ . 
\end{equation}

%%% Local Variables: 
%%% mode: latex
%%% TeX-master: "article"
%%% End: 

% vim: tw=70:spell

\section{Grad Moment System}
In this section, we derive the moment system of the Wigner equation
using the Grad type moment expansion.

\subsection{Hermite expansion of the distribution function} 
\label{sec:discretization}
Following the method in \cite{NRxx, NRxx_new}, we expand the
distribution function into Hermite series as
\begin{equation}
  \label{eq:expansion}
  f(t,\bx, \bv) = \sum_{\alpha \in \bbN^3}f_{\alpha}(t,\bx)
	\mH_{\mT,\alpha}
	\left( \frac{\bv-\bu(t,\bx)}{\sqrt{\mT(t,\bx)}} \right), 
\end{equation}
where $\alpha=(\alpha_1,\alpha_2,\alpha_3)$ is a three-dimensional
multi-index. The basis functions $\mH_{\mT,\alpha}$
are the 3-dimensional Hermite functions defined by
\begin{equation}
  \label{eq:base}
  \mH_{\mT,\alpha}(\bxi) =\prod\limits_{d=1}^3
 \frac{1}{\sqrt{2\pi}}\mT^{-\frac{\alpha_d+1}{2}}
\He_{\alpha_d}(\xi_d)\exp \left(-\frac{\xi_d^2}{2} \right),
\end{equation}
where $\He_{n}(x)$ is the Hermite polynomial of order $n$
\begin{equation}
  \label{eq:hermit}
  \He_n(x) = (-1)^n\exp \left( \frac{x^2}{2} \right) \frac{\dd^n}{\dd x^n} 
  \exp \left(-\frac{x^2}{2} \right).
\end{equation}
For convenience, $\He_{n}(x)$ is taken as zero if $n<0$, thus
$\mH_{\mT,\alpha}(\bxi)$ is zero when any component of
$\alpha$ is negative. The parameter $\mT$ in the expansion is
the scaled local temperature as
\begin{equation}
\label{eq:DefScaledTem}
\mT(t,\bx) = k_B T(t,\bx).
\end{equation}

It is clear that the equilibrium distribution $f_{\mathrm{eq}}$ is
coincidently equal to the first term of expansion, i.e.,
\begin{equation}
  f_{\mathrm{eq}}(t,\bx,\bv) = f_0(t,\bx) \mH_{\mT,0}
  \left( \frac{\bv-\bu(t,\bx)}{\sqrt{\mT(t,\bx)}}\right),
\end{equation}  
where $f_0(t,\bx) = \rho(t,\bx)$. So in this case the relaxation time
approximation scattering term \eqref{eq:collision} can be written as
the linear combination of the basis functions of orders greater than
or equal to $1$,
\begin{equation}
  \label{eq:collision}
  \frac{1}{\tau}(f_M-f) = -\frac{1}{\tau}\sum_{|\alpha| \geq 1}f_{\alpha}
  \mH_{\mT,\alpha} \left( \frac{\bv-\bu(t,\bx)}
  {\sqrt{\mT(t,\bx)}}\right).
\end{equation}

The definition of the Hermite function \eqref{eq:base} shows that 
each basis function is an exponentially decaying function multiplied 
by a multi-dimensional Hermite polynomial shifted by the local 
macroscopic momentum $\bu$ and scaled by the square root of the 
local temperature $\mT$. 

If one uses arbitrary known function $\bu'(t,\bx)$ and
$\mT'(t,\bx)$ in \eqref{eq:expansion} to expand the
distribution function $f(t,\bx,\bv)$ as
\begin{equation}
  \label{eq:expansion-o}
  f(t,\bx, \bv) = \sum_{\alpha \in \bbN^3}f_{\alpha}'(t,\bx)
  \mH_{\mT',\alpha}
  \left( \frac{\bv-\bu'(t,\bx)}{\sqrt{\mT'(t,\bx)}} \right), 
\end{equation}
then the following relations between the macroscopic quantities $\bu$,
$\mT$, the coefficients $f_{\alpha}$, and $\bu'$,
$\mT'$, the coefficients $f_{\alpha}'$ can be derived as
follows,
\begin{subequations}
\label{eq:similar_relation} 
 \begin{align}
    \rho &=  f_0 = f_0', \\
\rho \bu &=  \rho \bu' + (f_{e_d}')_{d=1,2,3}^T, \\
\rho |\bu - \bu'|^2 + 3\rho \mT &=  
\sum\limits_{d=1}^3(\mT' f_0' + 2 f_{2e_d}'),
\end{align}
\end{subequations}
where $e_d$ is the unit vector with its $d$-th entry to be $1$. It is
clear that the coefficients $f_\alpha$ expanded using parameters $\bu$
and $\mT$ satisfy the following conditions:
\begin{equation}
\label{eq:estimate}
  f_{e_i} = 0, \quad
  \sum\limits_{d=1}^3 f_{2e_d} = 0, \quad i =1, 2 , 3.
\end{equation} 
Moreover, if we define the heat flux $q_i$ and the pressure tensor
$P = \{p_{ij}\}, \ i,j =1,2,3$ with  
\begin{gather}
\label{eq:Qi1}
  q_i  = \frac{1}{2}\int_{\bbR^3}|\bv - \bu|^2(p_i - u_i)f
  \dd \bv,\\
\label{eq:Pij}
p_{ij} = \int_{\bbR^3}(p_i - u_i)(p_j - u_j)f \dd \bv , 
\end{gather}
then direct calculations give us the relations between them and  the
coefficients  $f_{\alpha}$ in \eqref{eq:expansion} as 
\begin{gather}
\label{eq:Qi2}
q_{i} = 2f_{3e_i} + \sum_{d=1}^3f_{2e_d+e_i}, \\ \label{eq:q_p}
p_{ij} - \frac{1}{3} \delta_{ij} \sum_{d=1}^3p_{dd} =
(1 + \delta_{ij}) f_{e_i+e_j}. 
\end{gather} 
By the definition of the temperature \eqref{eq:DefTemperature} and
\eqref{eq:DefScaledTem} and the definition of the tensor pressure
\eqref{eq:q_p}, the scaled temperature $\mT$ is a linear
combination of $p_{ij}$ as
\begin{equation}
\rho \mT = \dfrac{1}{3}\sum_{d=1}^{3} p_{dd}.
\end{equation}
With the relation \eqref{eq:q_p}, we then have
\begin{equation} \label{eq:pij_fij}
p_{ij} = \delta_{ij} \rho \mT +
(1 + \delta_{ij}) f_{e_i+e_j}. 
\end{equation}

\subsection{Moment expansion of the Wigner equation}
\label{moment_equations}
Now we are ready to derive the moment system by taking the moments of
the Wigner equation. The general method to get the moment system is to
first multiply the Wigner equation \eqref{eq:vlasov} of by polynomials
of momentum $\bp$ of different order and then integrate both sides
over momentum $\bp$ on $\bbR^3$. One equivalent way is as
follows. First, we substitute the expansion of the Wigner function
\eqref{eq:expansion} into the Wigner equation \eqref{eq:vlasov}, then
we collect the coefficients of basis functions of the same order at
both sides, and finally we equate the coefficients of the basis
functions of the same order on both sides to yield the derived moment
system.  We plug the Hermite series \eqref{eq:expansion} into the
Wigner equation \eqref{eq:vlasov}, and make calculations by noting
that the Hermite function \eqref{eq:base} used in this paper depends
also on the time $t$ and position $\bx$ through $\bu(t,\bx)$ and
$\mT(t,\bx)$, which is different from the general expansion
using the Hermite functions depending only on the momentum $\bv$
\cite{Shan1998}. For convenience, we list some useful relations of
Hermite polynomials as below \cite {Abramowitz}:
\begin{enumerate}
\item Orthogonality: $\displaystyle \int_{\bbR} \He_l(x)
\He_n(x) \exp(-x^2/2) \dd x = l! \sqrt{2\pi} \delta_{l,n}$;
\item Recursion relation: $\He_{n+1}(x) = x \He_n(x) - n \He_{n-1}(x)$;
\item Differential relation: $\He_n'(x) = n \He_{n-1}(x)$.
\end{enumerate}
And the following equality can be derived from the last two relations:
\begin{equation}
[\He_n(x) \exp(-x^2/2)]' = -\He_{n+1}(x) \exp(-x^2/2).
\end{equation}
Especially, we have
\begin{equation}\label{eq:partial_H}
  \pd{}{p_j} {\mH_{\mT,\alpha}\left( \frac{\bv-\bu}{\sqrt{\mT}} \right)}= 
  -\mH_{\mT,\alpha+e_j}\left(
    \frac{\bv-\bu}{\sqrt{\mT}} \right).
\end{equation}
With these relations, the part
\[
\pd{f}{t} + \bv \cdot \nabla_{\bx} f
\]
of \eqref{eq:vlasov} is expanded as
\begin{equation}
\label{eq:DriftExpansion}
\begin{split}
&~~\sum_{\alpha \in \bbN^3} \Bigg\{ \left(
    \frac{\partial f_{\alpha}}{\partial t} +
    \sum_{d=1}^3 \frac{\partial u_d}{\partial t} f_{\alpha-e_d} +
    \frac{1}{2} \frac{\partial \mT}{\partial t}
      \sum_{d=1}^3 f_{\alpha-2e_d}
  \right) \\
& \qquad + \sum_{j=1}^3 \Bigg[ \left(
    \mT \frac{\partial f_{\alpha - e_j}}{\partial x_j} +
    u_j \frac{\partial f_{\alpha}}{\partial x_j} +
    (\alpha_j + 1) \frac{\partial f_{\alpha+e_j}}{\partial x_j}
  \right) \\
& \qquad \qquad + \sum_{d=1}^3 \frac{\partial u_d}{\partial x_j}
    \left( \mT f_{\alpha-e_d-e_j} + u_j f_{\alpha-e_d}
      + (\alpha_j + 1) f_{\alpha-e_d+e_j} \right) \\
& \qquad \qquad + \frac{1}{2} \frac{\partial \mT}{\partial x_j}
    \sum_{d=1}^3 \left(
      \mT f_{\alpha-2e_d-e_j} + u_j f_{\alpha-2e_d}
      + (\alpha_j + 1) f_{\alpha-2e_d+e_j}
    \right)
  \Bigg] \Bigg\} \mH_{\mT,\alpha} \left(
    \frac{\bp - \bu}{\sqrt{\mT}}
  \right).
\end{split}
\end{equation}
Then using \eqref{eq:partial_H}, we calculate the pseudo-operator term
$\Theta[V]f$ expressed in \eqref{eq:ThetaVDiff}, and obtain 
\begin{equation}
\label{eq:ThetaVDiffExpansion}
(\Theta[V]f)(t,\bx,\bv)= 
 \sum_{\alpha,\blambda}
\frac{ (\hbar/2i) ^{ |\blambda| -1 }}{ \blambda ! }
\frac{ \partial ^{\blambda } V} 
{\partial \bx ^ {\blambda} } 
f_{\alpha - \blambda } 
\mH_{\mT,\alpha} \left(\frac{\bp - \bu}
		{\sqrt{\mT}} \right) ,
\end{equation} 
where the summation over $\blambda$ means the same as that in
\eqref{eq:ThetaVDiff}, and $f_{\alpha-\blambda}$ is taken as zero when
any component of $\alpha-\blambda$ is negative. Finally, the
scattering term $\left. \pd{f}{t} \right \vert_{\mathrm{Scat}}$
given in \eqref{eq:Maxwellian} becomes 
\begin{equation}
\label{eq:CollisionExpansion}
-\frac{1}{\tau}\sum_{|\alpha|\geq 2}f_{\alpha}\mH_{\mT,\alpha} \left(
    \frac{\bp - \bu}{\sqrt{\mT}}
  \right),
\end{equation}
noticing that $f_{e_d} = 0$, $d = 1, 2, 3$.

Collecting the three terms \eqref{eq:DriftExpansion},
\eqref{eq:ThetaVDiffExpansion} and \eqref{eq:CollisionExpansion}
yielded after the substitution of the Hermite expansion
\eqref{eq:expansion} into the Wigner equation \eqref{eq:vlasov}, we
can get the following general moment equations with a slight
rearrangement by matching the coefficients of the same weight
function:
\begin{equation} \label{eq:mnt_eq}
\begin{split}
& \frac{\partial f_{\alpha}}{\partial t} +
  \sum_{d=1}^3 \left(
    \frac{\partial u_d}{\partial t} +
    \sum_{j=1}^3 u_j \frac{\partial u_d}{\partial x_j} 
  \right) f_{\alpha-e_d}  
    +  \frac{1}{2} \left(
    \frac{\partial \mT}{\partial t} +
    \sum_{j=1}^3 u_j \frac{\partial \mT}{\partial x_j}
  \right) \sum_{d=1}^3 f_{\alpha-2e_d} \\
& \qquad\qquad + \sum_{j,d=1}^3 \Big[
    \frac{\partial u_d}{\partial x_j} \left(
      \mT f_{\alpha-e_d-e_j} + (\alpha_j + 1) f_{\alpha-e_d+e_j}
    \right) \\
&\qquad\qquad+ \frac{1}{2} \frac{\partial \mT}{\partial x_j} \left(
      \mT f_{\alpha-2e_d-e_j} + (\alpha_j + 1) f_{\alpha-2e_d+e_j}
    \right)
  \Big] \\
& \qquad \qquad+ \sum_{j=1}^3 \left(
    \mT \frac{\partial f_{\alpha - e_j}}{\partial x_j} +
    u_j \frac{\partial f_{\alpha}}{\partial x_j} +
    (\alpha_j + 1) \frac{\partial f_{\alpha+e_j}}{\partial x_j}
  \right)\\
&\qquad \qquad= -\frac{1}{\tau}
 \mathrm{H}( |\alpha|-2 )f_{\alpha}
 -\sum_{\blambda} 
 \frac{ (\hbar/2i) ^{ |\blambda| -1 }}{ \blambda ! }
 \frac{ \partial ^{\blambda } V} 
{\partial \bx ^ {\blambda} } 
f_{\alpha - \blambda } ,
\end{split}
\end{equation}
where $\mathrm{H}(x)$ is the Heaviside function defined by 
\begin{equation}
\label{eq:delta}
  \mathrm{H}(x) = \left\{
    \begin{array}{ll}
      0, & \text{if} ~~  x < 0, \\
      1, & \text{if}~~  x \geqslant 0.
    \end{array}
\right.
\end{equation} 
By setting $\alpha=0$ in \eqref{eq:mnt_eq}, we deduce the mass
conservation
\begin{equation}
\label{mass_con}
\pd{\rho}{t} +   
  \sum_{j=1}^3 \left(
    u_j \pd{\rho}{x_j} +
    \rho \pd{u_j}{x_j}
  \right) = 0.
 \end{equation}
By setting $\alpha = e_d$, with $ d = 1,2,3$ and 
noting that $f_{e_d} = 0$ in \eqref{eq:mnt_eq}, we obtain 
\begin{equation}
\label{eq:alpha = e_d}
\rho \left(
  \pd{u_d}{t} +
  \sum_{j=1}^3 u_j \pd{u_d}{x_j} 
\right) + \rho \pd{\mT}{x_d} 
  + \mT \pd{\rho}{x_d}
  + \sum_{j=1}^3 (\delta_{jd} + 1)
    \pd{f_{e_d + e_j}}{x_j} = -\rho \pd{V}{x_d},
\end{equation}
which is simplified as
\begin{equation} \label{eq:mtm}
\rho \left(
  \pd{u_d}{t} +
  \sum_{j=1}^3 u_j \pd{u_d}{x_j} 
\right) + \sum_{j=1}^3 \pd{p_{jd}}{x_j} = -\rho \pd{V}{x_d}.
\end{equation}
By setting $\alpha = 2e_d$, with $d=1,2,3$ and noting that 
$f_{e_d}=0$, we obtain
\begin{equation}
\label{eq:alpha=2e_d}
\begin{split}
\pd{f_{2e_d}}{t}&+\frac{\rho}{2}\left(\pd{\mT}{t}
            +\sum_{j=1}^3u_j\pd{\mT}{x_j}\right)
    +\rho\mT\pd{u_d}{x_d}
    +\sum_{j,l}(1+\alpha_j)f_{2e_d-e_l+e_j}\pd{u_l}{x_j}\\
&\qquad\qquad\qquad\qquad
    +\sum_{j=1}^3u_j\pd{f_{2e_d}}{x_j}+(1+2\delta_{jd})\pd{f_{2e_d+e_j}}{x_j}
    =-\frac{1}{\tau}f_{2e_d}.
\end{split}
\end{equation}
Noting that $\sum_{d=1}^3f_{2e_d}=0$, we sum the upper equations over
$d$ to get
% 计算
\begin{equation} 
\label{eq:moment_eq}
\rho \left( \pd{\mT}{t}
  + \sum_{j=1}^3 u_j \pd{\mT}{x_j} \right)
  + \frac{2}{3} \sum_{j=1}^3 \left(
    \pd{q_j}{x_j} +
    \sum_{d=1}^3 p_{jd} \pd{u_d}{x_j}
  \right) = 0.
\end{equation}
\comment{\begin{remark}
  We notice that $|\bv-\bu|^2$ is a linear sum of the first three
  Hermite polynomials and $f_{e_d}$, $d = 1, 2, 3$ are zero, thus
\begin{equation}
  \int_{\bbR^3} |\bv-\bu|^2 \pd{^{\blambda} f}{p^{\blambda}}\dd \bv 
  = 0, \quad |\blambda| \text{ is odd.}
\end{equation} 
The contribution of the nonlocal Wigner potential term does not appear
in \eqref{eq:moment_eq}.
\end{remark}}

Since $\rho\mT=\frac{1}{3}\sum_{d=1}^3p_{dd}$, we have
\begin{equation} \label{eq:partial_T2p}
\pd{\mT}{x_j} =
\frac{1}{3\rho}\sum_{d=1}^3\pd{p_{dd}}{x_j}-\frac{\mT}{\rho}\pd{\rho}{x_j},\quad
j=1,2,3.
\end{equation}
Substituting \eqref{eq:mtm}, \eqref{eq:moment_eq} and
\eqref{eq:partial_T2p} into \eqref{eq:mnt_eq}, we eliminate the time
derivatives of $\bu$ and $\mT$ and the spatial derivatives of
$\mT$. Then the quasi-linear form of
the moment system reads:
\begin{equation} \label{eq:mnt_system}
\begin{split}
&\pd{f_{\alpha}}{t}+\sum_{j=1}^3\left(
    \mT\pd{f_{\alpha-e_j}}{x_j}+u_j\pd{f_{\alpha}}{x_j}
    +(\alpha_j+1)\pd{f_{\alpha+e_j}}{x_j}\right) \\
&\qquad +\sum_{j=1}^3\sum_{d=1}^3\pd{u_d}{x_j}\left(
    \mT f_{\alpha-e_d-e_j} + (\alpha_j+1)f_{\alpha-e_d+e_j}
        -\frac{p_{jd}}{3\rho}\sum_{k=1}^3f_{\alpha-2e_k}\right)\\
&\qquad -\sum_{j=1}^3\sum_{d=1}^3
        \frac{f_{\alpha-e_d}}{\rho}\pd{p_{jd}}{x_j}
        -\frac{1}{3\rho}\left(\sum_{k=1}^3f_{\alpha-2e_k}\right)\sum_{j=1}^3\pd{q_j}{x_j}\\
&\qquad+\sum_{j=1}^3\left(\left(-\frac{\mT}{2\rho}\pd{\rho}{x_j}
        +\frac{1}{6\rho}\sum_{d=1}^3
           \pd{p_{dd}}{x_j}\right)\sum_{k=1}^3\left(\mT
                f_{\alpha-2e_k-e_j}+(\alpha_j+1)f_{\alpha-2e_k+e_j}\right)\right)\\
&\qquad=-\frac{1}{\tau}\mathrm{H}(|\alpha|-2)f_{\alpha}
    - \sum_{\blambda} 
    \frac{ (\hbar/2i) ^{ |\blambda| -1 }}{ \blambda ! }
    \pd{^{\blambda } V} 
    {\bx ^ {\blambda} } 
    f_{\alpha - \blambda } , \qquad \forall ~|\alpha| \geq 2.
\end{split}
\end{equation}
%\begin{equation} \label{eq:mnt_system}
%\begin{split}
%& \pd{f_{\alpha}}{t} - \frac{1}{f_0}
%  \sum_{d=1}^3 \sum_{j=1}^3
%    \pd{p_{jd}}{x_j} f_{\alpha-e_d}
%  - \frac{1}{3f_0} \sum_{j=1}^3 \left(
%    \pd{q_j}{x_j} +
%    \sum_{d=1}^3 p_{jd} \pd{u_d}{x_j}
%  \right) \sum_{d=1}^3 f_{\alpha-2e_d} \\
%& \quad + \sum_{j,d=1}^3 \left[
%    \pd{u_d}{x_j} \left(
%      \mT f_{\alpha-e_d-e_j} + (\alpha_j + 1) f_{\alpha-e_d+e_j}
%    \right) + \frac{1}{2} \pd{\mT}{x_j} \left(
%      \mT f_{\alpha-2e_d-e_j} + (\alpha_j + 1) f_{\alpha-2e_d+e_j}
%    \right)
%  \right] \\
%& \quad + \sum_{j=1}^3 \left(
%    \mT \pd{f_{\alpha - e_j}}{x_j} +
%    u_j \pd{f_{\alpha}}{x_j}\right) +
%    \sum_{j=1}^3 (\alpha_j + 1) \pd{
%        f_{\alpha+e_j}}{x_j}\\
%&   \quad=-\frac{1}{\tau}\mathrm{H}(|\alpha|-2)f_{\alpha}
%    - \sum_{|\blambda|>1} 
%    \frac{ (\hbar/2i) ^{ |\blambda| -1 }}{ \blambda ! }
%    \pd{^{\blambda } V} 
%    {\bx ^ {\blambda} } 
%    f_{\alpha - \blambda } , \quad \forall ~|\alpha| \geq 2.
%\end{split}
%\end{equation}
We should point out again the summation over $\blambda$ is extended
over all the non-negative integers $\lambda_d$, $d=1, 2, 3$ for which
$\vert \blambda \vert $ is odd and greater than $1$. Especially for
$|\alpha|\leq 2 $, the moment system \eqref{eq:mnt_system} derived
from the Wigner equation is the same as that derived from the
Boltzmann equation \cite{Li}.

With \eqref{eq:pij_fij}, we can have the equations for $p_{ij}$ by
\eqref{eq:mnt_system}. Precisely, we have the equation for $p_{ii}/2$,
$i = 1, 2, 3$, as
\begin{equation}\label{eq:moment_pi}
\begin{split}
 \pd{p_{ii}/2}{t} &+ \sum_{j=1}^3 u_j \pd{p_{ii}/2}{x_j}
  +\sum_{j=1}^3 \left( \frac{1}{2}+\delta_{ij} \right)
  \rho\mT\pd{u_j}{x_j} 
 + \sum_{j=1}^3\sum_{d=1}^3
  (2\delta_{ij}+1)f_{2e_i-e_d+e_j}\pd{u_d}{x_j} \\
&\qquad\qquad+\sum_{j=1}^3(2\delta_{ij}+1)\pd{f_{2e_i+e_j}}{x_j} = 
  -\frac{1}{2\tau}\left(p_{ii}-\frac{1}{3}\sum_{d=1}^3p_{dd}\right),
  \quad i = 1, 2, 3.
\end{split}
\end{equation}
If $i \neq j$, we have $p_{ij} = f_{e_i + e_j}$, thus its equation is
already in \eqref{eq:mnt_system}.

We collect the equations \eqref{mass_con}, \eqref{eq:mtm},
\eqref{eq:moment_pi} and \eqref{eq:mnt_system} together to obtain a
moment system with infinite number of equations. Noting that the
relation between $u_d$ and $f_{e_d}$ given in
\eqref{eq:similar_relation} and the definition of $q_i$ and of
$p_{ij}$ given in \eqref{eq:q_p}, we can see that what we obtain is a
quasi-linear system for $f_{\alpha}$. We would like to point out that
the only difference between the system derived from the Wigner
equation and that from the Boltzmann equation is the term with
high-order derivatives of the potential $V(t,\bx)$, which is a source
term of the quasi-linear system of $f_{\alpha}$.

%%% Local Variables: 
%%% mode: latex
%%% TeX-master: "article"
%%% End: 

% vim: tw=70:spell

\section{Moment Closure with Global Hyperbolicity}
\label{the moment closure}
The moment system derived from the Wigner equation consists of
\eqref{mass_con}, \eqref{eq:mtm}, \eqref{eq:moment_pi} and
\eqref{eq:mnt_system}. It is clear that this is a system with infinite
number of equations taken $\rho$, $u_d$, $p_{ij}$ and $f_\alpha$,
$|\alpha| \geqslant 3$, as unknowns. To obtain a system with finite
unknowns, we will truncate the expansion \eqref{eq:expansion} and
close the system following the method in \cite{Fan_new}.

With a truncation of \eqref{eq:expansion}, \eqref{eq:mnt_system} will
result in a finite moment system. Precisely, we let $M\geqslant 3$ be
a positive integer and only the coefficients in the set $\mathcal{M}
= \{f_{\alpha}\}_{|\alpha|\leqslant M}$ are considered. Let
$F_{M}(\bu, \mathcal{T})$ denotes the linear space spanned by all
$\mathcal{H}_{\mathcal{T}, \alpha}
\left(\frac{\bv -\bu(t,\bx)}{\sqrt{\mathcal{T}(t,\bx)}}\right)$'s
with $|\alpha|\leqslant M$,
and the expansion \eqref{eq:expansion} is truncated as
\begin{equation}
  \label{eq:expansion_1}
  f(t,\bx,\bv) \approx \sum_{|\alpha| \leqslant M} f_{\alpha}(t,\bx) \mathcal{H}_{\mathcal{T}, 
    \alpha}\left(\frac{\bv -\bu(t,\bx)}{\sqrt{\mathcal{T}(t,\bx)}}\right),
\end{equation}
with $f(t,\bx,\bv) \in F_{M}(\bu, \mathcal{T}) $ and $f_{\alpha} \in
\mathcal{M}$. The moment equations which contain
$\partial{f_{\alpha}}/\partial{t}$ with $|\alpha|>M$ are disregarded
in \eqref{eq:mnt_system}.  Then, \eqref{mass_con}, \eqref{eq:mtm},
\eqref{eq:moment_pi}  and \eqref{eq:mnt_system} with $2\leqslant
|\alpha|\leqslant M$ lead to a system with finite number of equations.

Following \cite{Fan_new}, we let
\[
\cS_M = \{\alpha\in\bbN^3\mid |\alpha| \le M\}.
\]
Then for any $\alpha\in \cS_M$, let
\begin{equation}\label{eq:def_mN}
\mN(\alpha) = \sum_{i=1}^3\binom{\sum_{k=4-i}^3\alpha_k + i-1}{i}+1
\end{equation}
to be the ordinal number of $\alpha$ in $\cS_M$, and the cardinal
number of set $\cS_M$ is \[N = \mN(M e_3) = \binom{M+3}{3},\] which is
total number of moments if a truncation with $|\alpha|\le M$ is
applied.

Let $\bw = (w_1, \cdots, w_N)^T \in\bbR^N$ and for each $i,j\in \{1,
2, 3\}$ and $i\neq j$,
\begin{subequations}\label{eq:basic_moments}
\begin{align}
w_1 &= \rho,    &   w_{\mN(e_i)} &= u_i,\\
w_{\mN(2e_i)} &= \frac{p_{ii}}{2},    &
    w_{\mN(e_i+e_j)} &= p_{ij}, \\
w_{\mN(\alpha)} &= f_{\alpha}, \quad 3\le|\alpha|\le M.
\end{align}
\end{subequations}
The moment system \eqref{mass_con}, \eqref{eq:mtm},
\eqref{eq:moment_pi} and \eqref{eq:mnt_system} is collected in
quasi-linear format as
\begin{equation}\label{eq:grad_system}
  \pd{\bw}{t} + \sum_{j = 1}^3 {\bf M}_j(\bw) \pd{\bw}{x_j} = 
  {\bf G} \bw,
\end{equation}
by taking the derivatives of $f_{\alpha + e_j}$, $|\alpha| = M$ to be
zero, where ${\bf M}_j$ and ${\bf G}$ are $N \times N$ matrices.  The
entries of ${\bf M}_j$ are given as the coefficients of the terms in
\eqref{mass_con}, \eqref{eq:mtm}, \eqref{eq:moment_pi} and
\eqref{eq:mnt_system} with derivatives of $\bw$. The entries of ${\bf G}$
arise from the nonlocal Wigner potential term and the scattering term.
From \eqref{eq:moment_pi}, one can observe that 
\begin{equation}\label{eq:matrix_pi}
{\bf G}_{\mN(2e_i),\mN(2e_j)} = -\frac{1}{\tau}
(\delta_{ij}-\frac{1}{3}), \quad i,j=1,2,3.
\end{equation}
And \eqref{eq:mnt_system} indicates that the diagonal entries of the
lower right part of ${\bf G}$ are
\begin{equation}\label{eq:matrix_diagonal}
  {\bf G}_{\mN(\alpha),\mN(\alpha)} = -\frac{1}{\tau}
  \mathrm{H}(|\alpha| - 2), \quad \text{ for } |\alpha|\ge 2
  \text{ and } \alpha\neq2e_i,~~i=1,2,3.
\end{equation}
From \eqref{eq:mtm}, it is clear that
\begin{equation} 
{\bf G}_{\mN(e_i),1} = -\dfrac{1}{\rho} \pd{V}{x_i}, \quad i = 1, 2, 3.
\end{equation}
The other nonzero entries of ${\bf G}$ from the nonlocal Wigner
potential are as
\begin{equation}\label{eq:G_a}
  {\bf G}_{\mN(\alpha), \mN(\alpha - \blambda)} = -\frac{ (\hbar/2i) 
  ^{ |\blambda| -1 }}{ \blambda ! } \frac{ \partial ^{\blambda } V} 
{\partial \bx ^ {\blambda} },
\end{equation}
where $|\blambda|$ is odd and $|\blambda|$ is greater than $1$ and
$|\alpha - \blambda| \neq 1$ or $2$. In case of $|\alpha - \blambda| =
1$, we have ${\bf G}_{\mN(\alpha), \mN(\alpha - \blambda)} = 0$ since
$f_{e_i} = 0$, $i = 1, 2, 3$. In case of $|\alpha - \blambda| = 2$,
we have \eqref{eq:G_a} if $\alpha - \blambda = e_i + e_j$, $i \neq j$,
since $w_{\mN(e_i + e_j)} = p_{ij} = f_{e_i + e_j}$. The difference is
in case of $\alpha - \blambda = 2 e_i$, $i = 1, 2, 3$. Precisely by
\eqref{eq:pij_fij}, we have
\begin{equation}
  {\bf G}_{\mN(\alpha),\mN(2 e_i)} = -\sum_{j=1}^3 \left(\delta_{ij} - 
    \dfrac{1}{3} \right) \frac{ (\hbar/2i)^{|\alpha| - 3}}{ (\alpha - 2 e_j) ! } 
  \frac{\partial^{\alpha - 2 e_j } V} {\partial \bx ^ {\alpha - 2 e_j} }.
\end{equation}

All other entries of ${\bf G}$ vanishes except for the ones specified
above. It is clear that the non-zero entries defined by
\eqref{eq:matrix_pi} and \eqref{eq:matrix_diagonal} by the scattering
term provide us a linear damping of the high order moment. The
contribution of the nonlocal Wigner potential matrix ${\bf G}$ is
strictly lower triangular. Thus these part of the matrix is nilponent.
As a result, the growth of the high order moments in time due to the
nonlocal Wigner potential term is essentially slower than exponential
growth rate.

In \eqref{eq:grad_system}, we following Grad \cite{Grad} take
${\partial f_{\alpha + e_j}}/{\partial x_j}$, $|\alpha| = M$, as zero
to make the system to be closed. It has been pointed out in
\cite{Fan_new} that it is not appropriate to set $\partial f_{\alpha +
  e_j}/\partial x_j = 0$, $|\alpha| = M$, as the closure proposed in
\cite{Grad} since the system is lack of hyperbolicity if the
distribution function is far away from the equilibrium. To obtain a
system with global hyperbolicity, we have to adopt the regularization
given in \cite{Fan_new}. For any $\alpha$ with $|\alpha| = M$, we
define
\begin{equation}\label{defRMj}
  \RM^j(\alpha) =
  (\alpha_j+1)\left[\sum_{d=1}^3 f_{\alpha-e_d+e_j}\pd{u_d}{x_j} + 
    \frac{1}{2}\left(\sum_{d=1}^3 f_{\alpha-2e_d+e_j}\right) 
  \pd{\mathcal{T}}{x_j}\right].
\end{equation}
and
\begin{equation} \label{eq:Grad_hme}
\hat{\bf M}_j \pd{\bw}{x_j} =
  {\bf M}_j \pd{\bw}{x_j} -
  \sum_{|\alpha|=M} \RM^j(\alpha) I_{\mN(\alpha)},
\quad \text{ for any admissible } \bw,
\end{equation}
where $I_k$ is the $k$-th column of the $N \times N$ identity
matrix. We regularize the system \eqref{eq:grad_system} as
\begin{equation}\label{eq:regularized_system}
  \pd{\bw}{t} + \sum_{j = 1}^3 \hat{\bf M}_j(\bw) \pd{\bw}{x_j} = 
  {\bf G} \bw ,
\end{equation}
which is the quantum hydrodynamics model we derived. It has been
proved in \cite{Fan_new} that
\begin{theorem} \label{thm:rot_inv} The regularized moment system
  \eqref{eq:regularized_system} is hyperbolic for any $\bw$ with
  positive temperature. Precisely, for a given unit vector $\bn =
  (n_1, n_2, n_3)$, the matrix
\begin{equation} \label{eq:rot_inv}
\sum_{j=1}^3 n_j \hat{\bf M}_j(\bw)
\end{equation}
is diagonalizable with eigenvalues as
\begin{equation}
\bu \cdot \bn + \rC{n}{m} \sqrt{\mathcal{T}},
  \qquad 1 \leqslant n \leqslant m \leqslant M+1,
\end{equation}
where $\rC{n}{m}$ is a root of $m$-order Hermite polynomial, and
satisfies $\rC{1}{m}<\cdots<\rC{m}{m}$. The structure of the $N$
eigenvectors can be fully clarified.
\end{theorem}
Based on this theorem, the regularized moment system
\eqref{eq:regularized_system} is locally well-posed due to the
hyperbolicity. We would like to mention here that the regularization
here actually does not add any new terms to the system
\eqref{eq:grad_system}. On the contrary, it has erased the terms in
\eqref{eq:mnt_system} with a factor $\alpha_j + 1$ in its coefficient
for the equations of $f_\alpha$ with $|\alpha| = M$ only.

%%% Local Variables: 
%%% mode: latex
%%% TeX-master: "article"
%%% End: 

%vim: tw=70:spell

\section{Regularized Moment System in 1D}
In 1D space, the structure of the system we derived is significantly
simpler than in multi-dimensional case. In this section, we give the
detailed formation of the moment system in 1D case since the 1D Wigner
equation is already very useful in modelling of RTDs
\cite{Frensley1987}.

The 1D Wigner equation is as
\begin{equation}
\label{eq:Wigner1D}
\pd{f}{t} + p \pd{f}{x} 
- \sum_{\lambda=1,3,5,\cdots} 
\frac{ (\hbar/2i)^{\lambda-1}}{\lambda !} 
\pd{^\lambda V}{x^\lambda} \pd{^\lambda f}{p^\lambda} = \frac{1}{\tau}
(f_{\mathrm{eq}} - f),
\end{equation}
where effective mass is again assumed to be $1$ for convinence.  The
equilibrium distribution $f_{\mathrm{eq}}$ is assumed to be a 1D
Maxwellian distribution
\begin{equation}
f_{\mathrm{eq}}(t,x,p) = \frac{\rho(t,x)}{
	\sqrt{2\pi k_B T} } \exp\left( - \frac{ ( p -u (t,x) )^2 }
		{2 k_B T(t,x) } \right). 
\end{equation}
The 1D Wigner function is expanded as
\begin{equation} \label{eq:expansion_1d}
f(t,x,p) = \sum_{n\in\bbN}f_n(t,x,p)
\mathcal{H}_{\mathcal{T},n}\left(  \frac{ p
		-u}{\sqrt{\mathcal{T}}}\right),
\end{equation} 
where 
\begin{equation}
\mathcal{H}_{\mathcal{T},n}(x) = 
\frac{1}{\sqrt{2\pi} } \mathcal{T}^{-(n+1)/2}
\He_{n}(x) \exp(-x^2/2 ), 
\end{equation}
and the 1D density $\rho$ and average momentum $u$ are defined by
\begin{equation}
\rho = \int_{-\infty}^{\infty} f(t,x,p) d p, \quad
u = \frac{1}{\rho}\int_{-\infty}^{\infty}p  f(t,x,p) d p,
\end{equation}
and the scaled temperature $\mathcal{T}$ defined by  
\begin{equation}
\mathcal{T} = k_B T = \frac{1}{\rho}
\int_{-\infty}^{\infty}(p -u)^2  f(t,x,p) d p.
\end{equation} 
The above equation plus the definition for the pressure \eqref{eq:Pij}
for $P$ shows that
\begin{equation}
\label{eq:P1D}
P = \rho \mathcal{T}.
\end{equation}
In one dimension, \eqref{eq:estimate} becomes  
\begin{equation}
f_1 = 0, \quad f_2 = 0. 
\end{equation} 
The relation \eqref{eq:Qi2} between the heat flux $q$ and the
coefficient $f_3$ of the expansion is turned to be
\begin{equation}
\label{eq:Q1D}
q = 3 f_3.
\end{equation} 

We present below the details of the regularized moment system
truncated at very low order based on \eqref{eq:regularized_system},
though the closed moment system up to any order can be written in a
general formation. We will demonstrate that with the low order
expansions of the distribution function, the moment system obtained is
able to capture very typical quantum effects. As the simplest case, a
closed regularized moment system truncated up to $M=3$ is
\begin{equation}
\label{eq:Moment0}
\pd{\rho}{t} + u \pd{\rho}{x} + \rho \pd{ u}{x} = 0, 
\end{equation}
\begin{equation}
\label{eq:Moment1}
\rho\pd{u}{t}+ \rho u \pd{u}{x}  + \pd{P}{x} = -\pd{V}{x}\rho,
\end{equation}
\begin{equation}
\label{eq:Moment2}
\pd{P/2}{t} + u \pd{P/2}{x} + \dfrac{3}{2} P 
\pd{u}{x} +  3\pd{f_3}{x} = 0.  
\end{equation}
\begin{equation}
\label{eq:Moment3}
 \pd{f_3}{t}
% +\dfrac{1}{2}\rho\mathcal{T}\pd{\mathcal{T}}{x} 
- \dfrac{P^2}{2\rho^2}\pd{\rho}{x} +\dfrac{P}{2\rho} \pd{P}{x} 
+u \pd{f_3}{x}  =
-\dfrac{1}{\tau} f_3 
+ \dfrac{\hbar^2}{24}\pd{^3 V}{x^3} \rho,
\end{equation}
where $\dfrac{\hbar^2}{24}\pd{^3 V}{x^3} \rho $ is the quantum
correction term yielded by the nonlocal Wigner potential. We
reformulate \eqref{eq:Moment0}, \eqref{eq:Moment1},
\eqref{eq:Moment2}, \eqref{eq:Moment3} into a matrix form
\begin{equation}
\label{eq:MomentMatrixForm}
\pd{\bw}{t} +{\bf M} (\bw) \pd{\bw}{x} = {\bf G} {\bw} , 
\end{equation} 
where $\bw=\left(\rho, u, P/2, f_3 \right)^{T}$ and 
\begin{equation} 
\label{eq:MatrixM3}
{\bf M} 
=\begin{pmatrix}
u & \rho & 0 & 0 \\
0 & u & \dfrac{2}{\rho} & 0 \\
0 & 3P & u & 3 \\
-\dfrac{P^2}{2\rho^2} & 0 & \dfrac{P}{\rho} & u 
\end{pmatrix},
\end{equation}
\begin{equation}
\label{eq:MatrixG3}
{\bf G} = 
\begin{pmatrix}
0 & 0 & 0 & 0 \\
-\pd{V}{x}\dfrac{1}{\rho} & 0 & 0 & 0 \\
0 & 0 & 0 & 0 \\
\dfrac{\hbar^2}{24}\pd{^3 V}{x^3} & 0 & 0 & -\dfrac{1}{\tau} 
\end{pmatrix}.
\end{equation} 

Here we consider the following example to show the difference between
the quantum moment system and the classical moment system. The
Boltzmann equation in 1D with a potential
\begin{equation}
V(x) =\left\{
\begin{array}[]{ll}
\exp \left( - \frac{1}{1-x^2} \right), & \text{if}\ \ |x|<1 , \\
0, & \text{else}, 
\end{array}
\right.
\end{equation}
admits a steady solution 
\begin{equation}
f(t,x,p) = \frac{1}{\sqrt{2\pi}} \exp\left( -\frac{p^2}{2} - V(x) \right), 
\end{equation}
with its moments as
\begin{equation}
\label{eq:RhoEx}
\rho = P = \exp(-V(x)), \quad u = f_3 = 0.
\end{equation}
$\rho$, $u$, $P$ and $f_3$ in \eqref{eq:RhoEx} satisfy the moment
system derived from the Boltzmann equation which can be obtained from
\eqref{eq:MomentMatrixForm} by removing the term $-\dfrac{\hbar^2}{24}
\pd{^3 V}{x^3} \rho$. The solutions given in \eqref{eq:RhoEx} show
that the classical behavior of particles obeys the Newtons law, i.e.,
the particles with low kinetic energy ($p^2/2$ at position $x$ ) can
not move to a new position with energy $V(x')$ which is greater than $
p^2/2 + V(x)$. As a result, $\rho$ in \eqref{eq:RhoEx} will not change
in time.

\begin{figure}[htbp]
\centering
\includegraphics[width=4in]{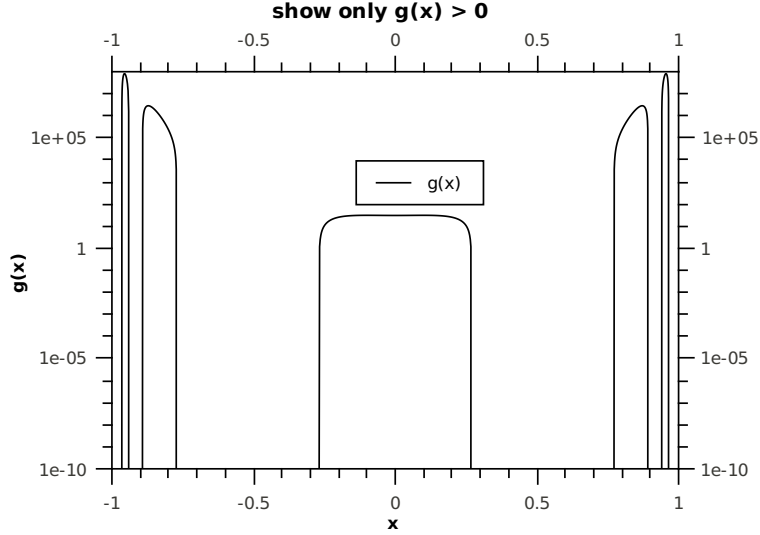}
\caption{The figure of $g(x)$ in \eqref{eq:gx}. The curve  of $g(x)$
is plotted in semi-log scale only in the subintervals where
	$g(x)>0$.  $g(x)=0$ for $x$ is outside of $(-1,1)$ 
	has not been plotted for it is obvious that $g(x)=0$.    }
\label{fig:gx}
\end{figure}

In the quantum mechanics, the particles behave like a wave that the
particles with lower energy can be tunneled through a potential
barrier. The moment system derived from the Wigner equation will
reflect this point. Let us consider the system
\eqref{eq:MomentMatrixForm} with \eqref{eq:RhoEx} as initial value. We
carry out an asymptotic analysis to study the behavior of the solution
around $t = 0$, using time $t$ as the asymptotic expansion
parameter. The ansatz we adopted is as
\begin{align}
\label{eq:f3Asympotic}
f_3 & = f_{3,0} + f_{3,1} t + \mathcal{O}(t^2), \\ 
\label{eq:PAsympotic}
P & = P_0 + P_1 t +  P_2 t^2 + \mathcal{O}(t^3), \\
\label{eq:uAsympotic}
u & = u_0 + u_1 t +  u_2 t^2 + u_3 t ^3 +  \mathcal{O}(t^4), \\ 
\label{eq:rhoAsympotic}
\rho & = \rho_0 + \rho_1 t + \rho_2 t^2 + \rho_3 t ^3 + \rho_4 t ^4 
+\mathcal{O}(t^5),
\end{align}
where the leading order terms are as the initial value
\begin{equation}
\label{eq:ZeroOrder}
\rho_0 = P_0 = \exp(-V(x)), \quad u_0 = f_{3,0} = 0.
\end{equation}
Substituting the ansatz into \eqref{eq:Moment3}, we immediately have
\begin{equation}
f_{3,1} = \frac{\hbar^2}{24} V^{(3)}(x)  \rho_0,  
\end{equation} 
thus $f_3$ is as
\begin{equation}
\label{eq:f3Result}
f_3 =  \frac{\hbar^2}{24} V^{\left( 3 \right)}(x) \rho_0 t +
\mathcal{O}(t^2)
\end{equation} 
Plugging \eqref{eq:f3Result} and \eqref{eq:PAsympotic} -
\eqref{eq:rhoAsympotic} into \eqref{eq:Moment2}, we obtain
\begin{equation}
P_1 = 0 , \quad P_2 =  -\frac{ \hbar^2}{24} \opd{}{x} \left(V^{\left( 3
			\right)}(x) \rho_0\right) , 
\end{equation}
which yields 
\begin{equation}
\label{eq:PResult}
P = P_0  - \frac{\hbar^2}{8} \opd{}{x}\left(V^{\left( 3 \right)}(x) 
		\rho_0 \right) t^2 + \mathcal{O}(t^3).
\end{equation}
Similar procedure gives us
\begin{equation}
\label{eq:uResult}
u = u_0 + \frac{\hbar^2}{24} \opd{^2}{x^2}\left(V^{\left( 3 \right)}(x) 
		\rho_0 \right)t ^3 + \mathcal{O}(t^4),
\end{equation}
\begin{equation}
\label{eq:rhoResult}
\rho = \rho_0 - \frac{\hbar^2}{96} \opd{^3 }{x^3 } \left ( 
		V^{(3)} \rho_0^2  \right) t ^4 +
\mathcal{O}(t^5).
\end{equation}
Noting that $\rho_0 = \exp(-V(x))$, we can rewrite
\eqref{eq:rhoResult} into
\begin{equation}
\label{eq:rhoResultAgain}
\rho = \rho_0 + \frac{\hbar^2 t^4 }{96} 
g(x) + \mathcal{O}(t^5), 
\end{equation}
where 
\begin{equation}
\label{eq:gx}
g(x)=	-\opd{^3 }{x^3 }\left ( V^{(3)} \exp(-2V) \right) ,  
\end{equation}
whose figure is plotted in \figref{fig:gx}.

%the following wxmaxima command gives the data 
%for \figref{fig:gx}, then qtiplot will output the eps figure
%load(draw);
%V(x):=exp(1/(x^2-1));
%g(x):= -diff(exp(-2*V(x))*diff(V(x),x,3),x,3);
%plot2d(g(x),[x,-0.999,0.999]);
%draw2d( terminal =eps, explicit(g(x), x, -0.999, 0.999)) ;
%
From \eqref{eq:f3Result}, we see that the heat flux $f_3$ is changed
by the term $\dfrac{\hbar^2}{24} \pd{^3 V}{x^3} \rho_0$, which is the
difference between the Boltzmann equation and the Wigner equation. The
change of heat flux induces the change of the temperature
\eqref{eq:PResult} thus the electrons in part of the domain become
cooler than the electrons in the else part of the domain. This makes
the electrons with higher temperature moving to the domain with lower
temperature \eqref{eq:uResult}, eventually resulting the
redistribution of the density of the electrons
\eqref{eq:rhoResult}. \eqref{eq:rhoResultAgain} shows that how the
density $\rho$ changes at a very small $t$, and whether it will
increase or decrease depends on the sign of $g(x)$, which is shown in
\figref{fig:gx}. So it is observed from \figref{fig:gx} that the
particles will be redistributed because of the quantum
term. Precisely, at a very small time $t$, the electron density in the
subintervals of $(-1, 1)$ where $g(x)>0$, whose curves are plotted in
\figref{fig:gx}, is increasing, and the electron density in the rest
subintervals of $(-1, 1)$ where $g(x)<0$ without its curves plotted,
is decreasing.

%We can also consider $V(x) = exp(-x^2/2)$, and this derivatives 
%is not very big and the zeros are clearer from its figure. 

In the remain part of this section, we present the 1D system obtained
by truncating the expansion \eqref{eq:expansion_1d} at $M = 4$, $5$
and $6$ for the reader's convenience to carry out numerical
simulations. If we truncate up to $M = 4$, we obtain a closed system
consisting of \eqref{eq:Moment0} - \eqref{eq:Moment2} and
\begin{equation}
\label{eq:Moment3_B}
\pd{f_3}{t} 
+ 4 f_3 \pd{u}{x}  
+ \dfrac{1}{2} \mathcal{T}\rho   \pd{\mathcal{T}}{x} 
+u \pd{f_3}{x} + 4 \pd{f_4}{x} 
= -\dfrac{f_3}{\tau}  
+ \dfrac{\hbar^2}{24}\pd{^3 V}{x^3} \rho,
\end{equation}
and
\begin{equation}
\label{eq:Moment4} 
\pd{f_4}{t} 
-\dfrac{f_3}{\rho} \pd{P}{x} 
+ \mathcal{T} \pd{f_3}{x} + u \pd{f_4}{x} 
= -\dfrac{f_4}{\tau}. 
\end{equation} 
And the system is collected into the matrix form
\eqref{eq:MomentMatrixForm} with $ \bw=\left(\rho, u, P/2, f_3,
  f_4\right)^{T}$ and
\begin{equation}
\label{eq:MatrixM4}
{\bf M} 
=\begin{pmatrix}
u & p & 0 & 0 & 0 \\
0 & u & \dfrac{2}{\rho} & 0 & 0  \\
0 & 3P & u & 3 & 0   \\
-\dfrac{P^2}{2\rho^2} & 4f_3 & \dfrac{P}{\rho} & u & 4 \\
0 & 0 & -\dfrac{f_3}{ \rho} & \dfrac{P}{\rho} & u 
\end{pmatrix},
\end{equation}
\begin{equation}
\label{eq:MatrixG4}
{\bf G} = 
\begin{pmatrix}
0 & 0 & 0 & 0 & 0  \\
-\pd{V}{x}\dfrac{1}{\rho} & 0 & 0 & 0 & 0  \\
0 & 0 & 0 & 0 & 0  \\
\dfrac{\hbar^2}{24}\pd{^3 V}{x^3} & 0 & 0 & -\dfrac{1}{\tau} & 0  \\
0 & 0 & 0 &  0  & -\dfrac{1}{\tau}   
\end{pmatrix}.
\end{equation} 

Similarly we can write out the system truncated at $M = 5$ which
consists of \eqref{eq:Moment0} - \eqref{eq:Moment2},
\eqref{eq:Moment3_B} and
\begin{equation}
\label{eq:Moment4_C} 
\pd{f_4}{t} 
-\dfrac{f_3}{\rho}  \pd{P}{x}  
+5 f_4 \pd{u}{x}  + \frac{5f_3}{2}  \pd{\mathcal{T}}{x} 
+ \mathcal{T} \pd{f_3}{x} + u \pd{f_4}{x} 
+ 5 \pd{f_5}{x} 
= -\dfrac{f_4}{\tau} ,
\end{equation} 
and
\begin{equation}
\label{eq:Moment5} 
\begin{split}
&
\pd{f_5}{t} 
-\dfrac{f_4}{\rho} \pd{P}{x}  
-\dfrac{3f_3}{\rho}\pd{f_3}{x} 
+ \mathcal{T} \pd{f_4}{x} + u \pd{f_5}{x} 
 = -\dfrac{f_5}{\tau} 
+ \dfrac{\hbar^4}{1920} \pd{^5 V}{x^5} f_0 , 
\end{split}
\end{equation} 
which is expressed into the matrix form 
\eqref{eq:MomentMatrixForm} with
where $\bw=\left(\rho, u, P/2, f_3, f_4, f_5\right)^{T}$ and 
\begin{equation}
\label{eq:MatrixM5}
{\bf M} 
=\begin{pmatrix}
u & p & 0 & 0 & 0  & 0 \\
0 & u & \dfrac{2}{\rho} & 0 & 0  & 0  \\
0 & 3P & u & 3 & 0 & 0    \\
-\dfrac{P^2}{2\rho^2} & 4f_3 & \dfrac{P}{\rho} & u & 4 & 0  \\
-\dfrac{5P f_3}{2\rho^2}  & 5 f_4 & \dfrac{3f_3}{\rho} & \dfrac{P}{\rho} & u & 5 \\   
0 & 0 & -\dfrac{ f_4}{\rho} & -\dfrac{3f_3}{\rho}
& \dfrac{P}{\rho} & u  
\end{pmatrix},
\end{equation}
\begin{equation}
\label{eq:MatrixG5}
{\bf G} = 
\begin{pmatrix}
0 & 0 & 0 & 0 & 0  & 0  \\
-\pd{V}{x}\dfrac{1}{\rho} & 0 & 0 & 0 & 0 & 0   \\
0 & 0 & 0 & 0 & 0  & 0  \\
\dfrac{\hbar^2}{24}\pd{^3 V}{x^3} & 0 & 0 & -\dfrac{1}{\tau} & 0  & 0 \\
0 & 0 & 0 &  0  & -\dfrac{1}{\tau} &  0 \\ 
-\dfrac{\hbar^4}{1920}\pd{^5 V}{x^5} & 0 & 0 & 0 & 0  
& -\dfrac{1}{\tau}   
\end{pmatrix}.
\end{equation} 

In the case of truncating at $M = 6$, the moment system obtained
consists of \eqref{eq:Moment0} - \eqref{eq:Moment2},
\eqref{eq:Moment3_B}, \eqref{eq:Moment4_C} and
\begin{equation}
\label{eq:Moment5_D} 
\begin{split}
\pd{f_5}{t} 
-\dfrac{f_4}{\rho} \pd{P}{x}  
&-\dfrac{3f_3}{\rho}\pd{f_3}{x} 
 +6 f_{5} \pd{u}{x}  
+ 3f_{4} \pd{\mathcal{T}}{x}  \\
 &+ \mathcal{T} \pd{f_4}{x} + u \pd{f_5}{x} 
+ 6 \pd{f_6}{x}  
= -\dfrac{f_5}{\tau} 
- \dfrac{\hbar^4}{1920} \pd{^5 V}{x^5}f_0 ,
\end{split}
\end{equation} 
\begin{equation}
\label{eq:Moment6} 
\begin{split}
&
\pd{f_6}{t} 
-\dfrac{f_5}{\rho} \pd{P}{x}  
-\dfrac{3f_4}{\rho}\pd{f_3}{x} 
+\dfrac{1}{2}\pd{\mathcal{T}}{x}  \mathcal{T} f_3 
%+ \pd{\mathcal{T}{x} f_{5} 
+ \mathcal{T} \pd{f_5}{x} + u \pd{f_6}{x} 
%截断去掉
%+ 2 \pd{f_6}{x}   
 = -\dfrac{f_6}{\tau} 
+ \dfrac{\hbar^2}{24}\pd{^3 V}{x^3} f_3 . 
\end{split}
\end{equation} 
This system can be formulated in the matrix form of
\eqref{eq:MomentMatrixForm} with $\bw=\left(\rho, u, P/2, f_3, f_4,
  f_5, f_6\right)^{T}$ and
\begin{equation}
\label{eq:MatrixM6}
{\bf M} 
=\begin{pmatrix}
u & p & 0 & 0 & 0  & 0 & 0 \\
0 & u & \dfrac{2}{\rho} & 0 & 0  & 0 & 0  \\
0 & 3P & u & 3 & 0 & 0  & 0   \\
-\dfrac{P^2}{2\rho^2} & 4f_3 & \dfrac{P}{\rho} & u & 4 & 0 & 0 \\
-\dfrac{5P f_3}{2\rho^2}  & 5 f_4 & \dfrac{3f_3}{2\rho} & \dfrac{P}{\rho} & u & 5 & 0 \\   
-\dfrac{3P f_4}{\rho^2}  & 6f_5 &  \dfrac{2f_4}{\rho} 
 & -\dfrac{3f_3}{\rho}
& \dfrac{P}{\rho} & u  & 6  \\
-\dfrac{P^2 f_3}{2 \rho^3}  & 0 & \dfrac{Pf_3}{2\rho^2}-\dfrac{f_5}{\rho} 
&  -\dfrac{3 f_4}{\rho} & 0 
& \dfrac{P}{\rho} & u  
\end{pmatrix},
\end{equation}
\begin{equation}
\label{eq:MatrixG6}
{\bf G} = 
\begin{pmatrix}
0 & 0 & 0 & 0 & 0  & 0 & 0   \\
-\pd{V}{x}\dfrac{1}{\rho} & 0 & 0 & 0 & 0 & 0 & 0    \\
0 & 0 & 0 & 0 & 0  & 0 & 0  \\
\dfrac{\hbar^2}{24}\pd{^3 V}{x^3} & 0 & 0 & -\dfrac{1}{\tau} & 0  & 0 &
0 \\
0 & 0 & 0 &  0  & -\dfrac{1}{\tau} & 0 &
  0 \\ 
-\dfrac{\hbar^4}{1920}\pd{^5 V}{x^5} & 0 & 0 & 0 & 0   & 
-\dfrac{1}{\tau} &  0 \\   
0 & 0 & 0 & \dfrac{\hbar^2}{24}\pd{^3 V}{x^3} & 0 & 
0 &
-\dfrac{1}{\tau}       
\end{pmatrix}.
\end{equation} 

Observing {\bf M}'s in \eqref{eq:MatrixM3}, \eqref{eq:MatrixM4},
\eqref{eq:MatrixM5}, \eqref{eq:MatrixM6} and {\bf G}'s in
\eqref{eq:MatrixG3}, \eqref{eq:MatrixG4}, \eqref{eq:MatrixG5},
\eqref{eq:MatrixG6}, we find that the left hand side of the moment
system of \eqref{eq:MomentMatrixForm} is the exactly the same as the
moment system derived from the classical Boltzmann equation, and the
quantum mechanical terms with a typical characteristic which involves
the high order derivatives of the potential $V$ are only appear in the
matrices $G$ in the right-hand side of \eqref{eq:MomentMatrixForm}. It
is clear that every $G$ is lower triangular. The formation of matrices
$G$ shows that the quantum potential term works in the way by letting
high-order moments get information from the moments of different lower
orders, as low as to order $0$.

%上面是我的感性说法，需要更多的数学讨论，说明量子效应是如何体现表现出来。

%%% Local Variables: 
%%% mode: latex
%%% TeX-master: "article"
%%% End: 

% vim: tw=70:spell

\section{Conclusion}
We extend the moment closure method \cite{Fan_new} for the Boltzmann
equation to its quantum counterpart, the Wigner equation. And we
obtain a hyperbolic moment system, i.e., quantum hydrodynamic system.
The method to derive the quantum hydrodynamic system with the global
hyperbolicity is systematic that systems for arbitrary number of
moments are obtained at once. Numerical simulations are to be carried
out to demonstrate that the quantum effects are able to be captured by
using the quantum hydrodynamic model derived.

\section*{Acknowledgements}
This research of R. Li was supported in part by the
National Basic Research Program of China (2011CB309704) and Fok Ying
Tong Education and NCET in China. T. Lu was supported in part by the 
NSFC (11011130029) and by SRF for ROCS, SEM.

%%% Local Variables: 
%%% mode: latex
%%% TeX-master: "article"
%%% End: 

\bibliographystyle{plain}
\bibliography{../article,../tiao}
\end{document}